\documentclass[3p]{elsarticle}
\usepackage{amssymb}
\usepackage{graphicx}
\usepackage{amsmath}

\renewcommand{\vec}[1]{\ensuremath\boldsymbol{#1}}

\title{A numerical model for the trans-membrane voltage of vesicles
	\tnoteref{th1}}
\tnotetext[th1]{This work was supported by the University at Buffalo SUNY and NSF Award \#1253739}
\author[ub]{EM.~Kolahdouz}
\author[ub]{D.~Salac\corref{cor1}}
\cortext[cor1]{Corresponding author}
\address[ub]{Department of Mechanical and Aerospace Engineering, University at Buffalo SUNY, Buffalo, NY, 14260.}

\begin{document}

	\begin{abstract}
	
	The Immersed Interface Method is employed to solve the time-varying electric field equations around
	a three-dimensional vesicle. To achieve second-order accuracy 
	the implicit jump conditions for the electric potential, up to the second normal derivative, are derived. 
	The trans-membrane potential is determined implicitly as part of the algorithm. 
	The method is compared to an analytic solution based on spherical harmonics and verifies the 
	second-order accuracy of the underlying discretization even in the presence of solution discontinuities.
	A sample result for an elliptic interface is also presented.
	
	\end{abstract}


	\maketitle

\section{Introduction}
\label{sec:1.0}


In this work a new numerical method is developed to obtain the time-varying electric and trans-membrane potentials 
associated with a lipid vesicle membrane exposed to electric fields. This work is part of a larger effort to understand the general electrohydrodynamics
of lipid vesicles.
The method presented here is robust and can be applied to any vesicle shape. 

\begin{figure}
	\centering	
	\begin{minipage}{0.4\textwidth}
		\centering
		\includegraphics[width=1.5in]{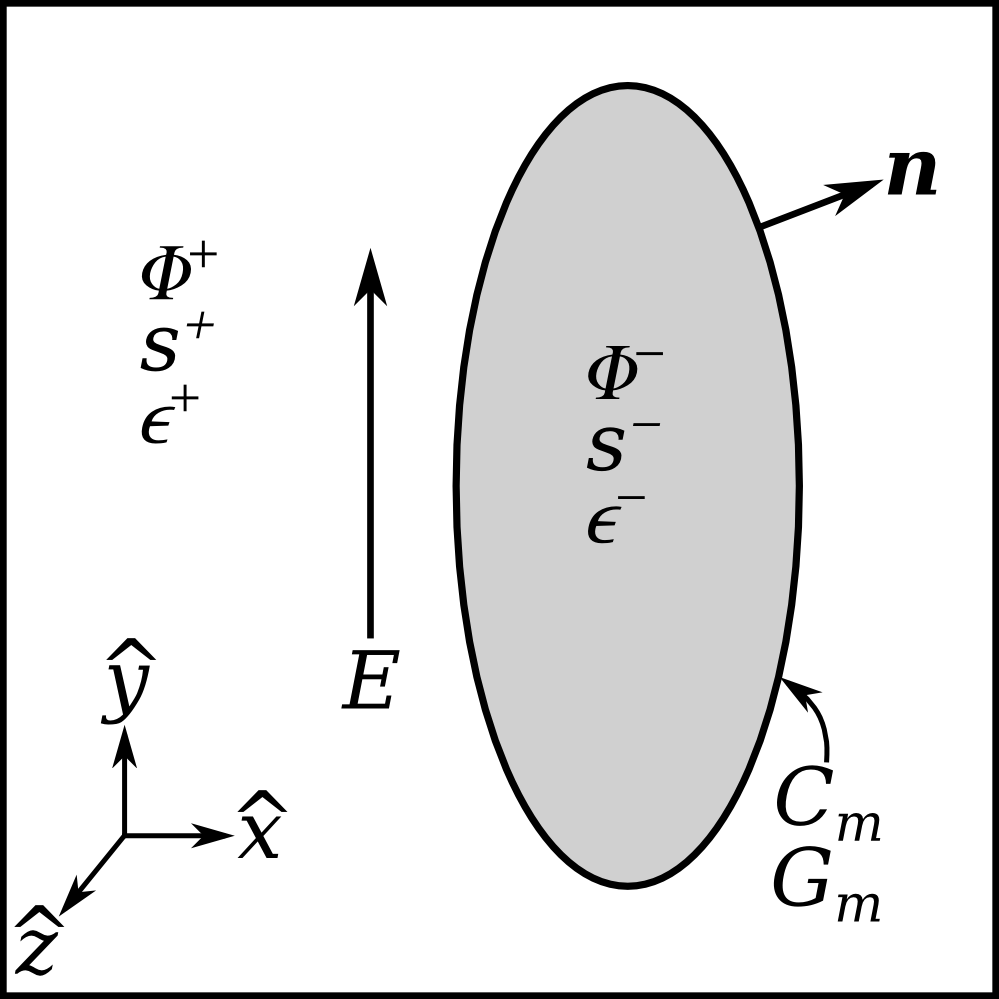}
		\caption{The system of interest: A vesicle exposed to an electric field. Properties differ between the inner and outer fluid.}
		\label{fig:sketchelec}		
	\end{minipage} \hspace{2cm}
	\begin{minipage}{0.4\textwidth}
		\centering
		\includegraphics[width=1.5in]{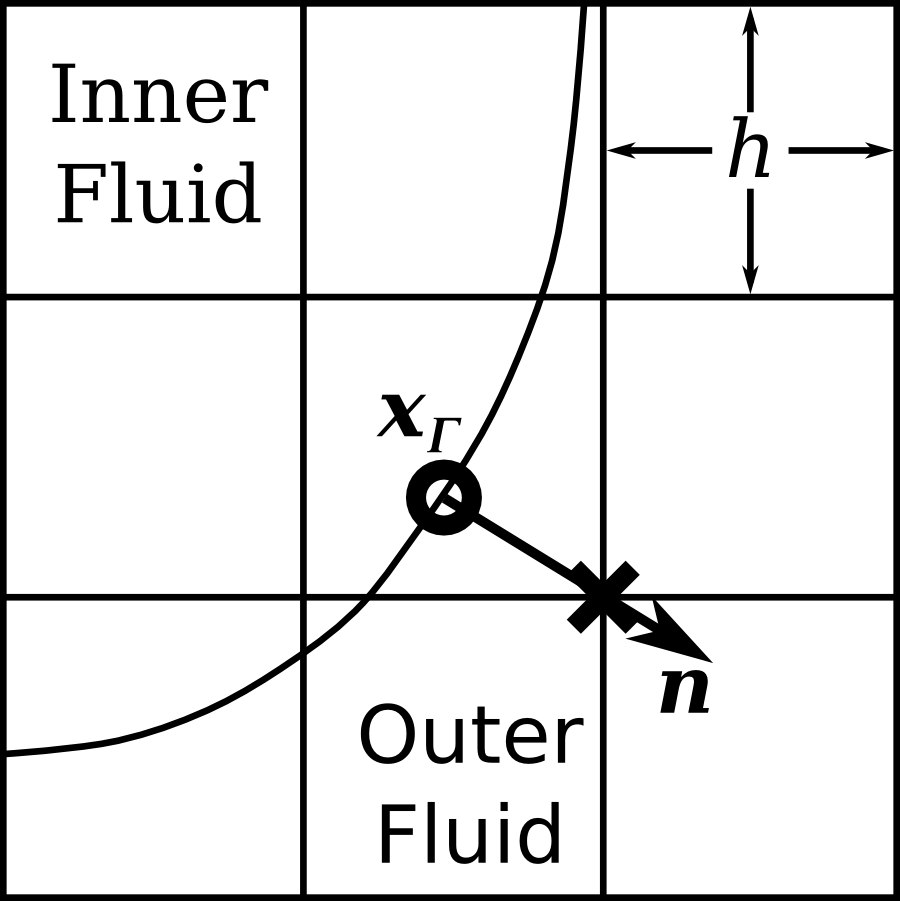}
		\caption{Sample grid showing interface and outward facing normal. The grid node
		denoted by the cross requires corrections due to discontinuities in the stencil.
		The corrections are calculated 
		at the circle and extended out to the grid node.}
		\label{fig:grid}		
	\end{minipage}	
\end{figure}

Consider a lipid bilayer vesicle exposed to an electric field, Fig. \ref{fig:sketchelec}. The vesicle is assumed to be made of a charge-free lipid bilayer
membrane with capacitance $C_m$ and conductivity $G_m$. It is suspended in an outer fluid denoted as $\Omega^+$ with conductivity $s^+$ and permittivity $\epsilon^+$. 
The inner fluid, denoted as $\Omega^-$, is enclosed by the vesicle and assumed to have a different conductivity $s^-$ and permittivity constant $\epsilon^-$. 

Application of an electric field causes a redistribution of bulk charge density in both inside and outside of the membrane \cite{Schwalbe2011,Melcher1969}.
Denote the electric potential in the inner fluid as $\Phi^-$ and in the outer fluid as $\Phi^+$. As there 
is no local free charge density in either fluid, the electric potential in each fluid is given as the solution to $\nabla^2\Phi^\pm=0$ \cite{Saville1997}.

When an electric field is applied to the system, charges will accumulate on both the inner and outer sides of membrane due 
to the ion impermeability of the lipid bilayer.
This turns the membrane into a capacitive interface, which results in a discontinuity of the electric potential across the domain \cite{Schwalbe2011},
\begin{equation}
	\left[\Phi\right]=\lim_{a\rightarrow 0}\Phi^{+}\left(\vec{x}_\Gamma+a\vec{n}\right)-\lim_{a\rightarrow 0}\Phi^{-}\left(\vec{x}_\Gamma-a\vec{n}\right)=-V_m(t),
	\label{eq:Phi_Difference}
\end{equation}
where $V_m(t)$ is the time-varying trans-membrane potential, $\vec{x}_\Gamma$ is a point on the membrane,
and $\vec{n}$ is the outward unit normal pointing into $\Omega^+$, see Fig. \ref{fig:grid}. For simplicity the
limit notation will be dropped henceforth.

The trans-membrane potential itself can be obtained from the conservation of current density across the membrane \cite{Seiwert2012,DeBruin1999},
\begin{equation}
	C_m\frac{dV_m}{dt}+G_{m}V_{m}=\vec{n}\cdot \bigl(s^+\vec{E}^+\bigl)=\vec{n}\cdot \bigl(s^-\vec{E}^-\bigl).
	\label{eq:Vm_equation}
\end{equation}
Assuming that the membrane conductance and capacitance have uniform and constant values on the interface, the trans-membrane 
potential will only depend on changes in the surrounding domain electric potential and the interface shape.

The bulk Ohmic current, $\vec{J}=s\vec{E}$, is continuous in the normal direction across the membrane. Therefore
$\vec{n}\cdot(\vec{J^{+}} - \vec{J^{-}})=\vec{n}\cdot(s^{+}\vec{E^{+}} - s^{-}\vec{E^{-}})=0$.	
However, there is a discontinuity in the normal component of displacement vector due to induced charges on the membrane
 $\vec{n}\cdot(\epsilon^{+}\vec{E^{+}} - \epsilon^{-}\vec{E^{-}})= Q$,	
where $Q$ is the induced charge density at the top or bottom of the membrane. This net charge imbalance occurs across the interface due to the difference in
physical and electrical properties of the inner and outer fluids. $Q$ is introduced here only for the sake of completeness and is not used in the calculations 
for the electric potential. 

\section{Electric Potential Jump Conditions}
\label{sec:2.0}

Let the electric potential field in the entire domain (inner plus outer fluid) be denoted as $\Phi$. 
Consider solving for the potential in the entire domain, $\nabla^2\Phi=0$, using a finite difference method.
Large errors are introduced into the solution near the interface due to the discontinuous electric potential field.
The Immersed Interface Method was first introduced by Leveque and Li to address the issue of solving discontinuous PDEs 
across an embedded interface \cite{LeVeque1994}.
To produce accurate solutions the jump 
of the solution across the interface are included in the numerical discretization.
This method has been used extensively to solve elliptic problems with interfaces \cite{LeVeque1994,Li1998} and later was extended to model the
Stokes or Navier-Stokes equations with singular forces and discontinuous viscosity \cite{Li2001,Le2006}. 
The IIM is also able to handle sharp interfaces with discontinuities and singularities in the coefficients and the solutions \cite{Li2003}. 

In this work a simplified Immersed Interface Method is used \cite{Lai2008}. To obtain second-order spatial accuracy in the solution the 
jumps in the electric potential and up to the second normal derivative are needed across the interface \cite{Lai2008}. The first jump condition
is obtained directly from the time-varying trans-membrane voltage, Eq. (\ref{eq:Phi_Difference}). The time-varying nature of this jump condition will
be handled in the next section.

To derive the jump condition for the first normal derivative of electric potential the  
continuity of current density across the interface is used:
\begin{align}
	0 &= s^+\frac{\partial\Phi^+}{\partial n}-s^-\frac{\partial\Phi^-}{\partial n}, \\
	0 &= s^+\frac{\partial\Phi^+}{\partial n}-s^-\frac{\partial\Phi^-}{\partial n} + s^-\frac{\partial\Phi^+}{\partial n} - s^-\frac{\partial\Phi^+}{\partial n}, \\
	0 &= \left(s^+-s^-\right)\frac{\partial\Phi^+}{\partial n}+s^-\left(\frac{\partial\Phi^+}{\partial n}-\frac{\partial\Phi^-}{\partial n}\right), \\
	0 &= \left[s\right]\frac{\partial\Phi^+}{\partial n}+s^-\left[\frac{\partial\Phi}{\partial n}\right].
\end{align}
Similarly it can be shown that 
\begin{equation}
	0 = \left[s\right]\frac{\partial\Phi^-}{\partial n}+s^+\left[\frac{\partial\Phi}{\partial n}\right].
\end{equation}
Solving for the jump in the normal electric field yields
\begin{equation}
	\biggl[\frac{\partial\Phi}{\partial n}\biggl]=-\frac{[s]}{s^{+}}\frac{\partial\Phi^{-}}{\partial n}=-\frac{[s]}{s^{-}}\frac{\partial\Phi^{+}}{\partial n}.
	\label{eq:Phi_Prime_Jump}
\end{equation}

For the jump in the second normal derivative, start with the relation between the Laplacian and the surface Laplacian of an arbitrary scalar function, 
$\nabla^2\Phi=\nabla^2_s\Phi+H\tfrac{\partial\Phi}{\partial n}+\tfrac{\partial^2\Phi}{\partial n^2}$,
where $\nabla^2_s=\left[\left(\vec{I}-\vec{n}\otimes\vec{n}\right)\nabla\right]\cdot\left[\left(\vec{I}-\vec{n}\otimes\vec{n}\right)\nabla\right]$ is the surface Laplacian and $H=\nabla\cdot\vec{n}$ is the summation of the two principle curvatures. 
Applying the jump operator results in $[\nabla^2\Phi]=[\nabla^2_s\Phi]+H[\tfrac{\partial\Phi}{\partial n}]+[\tfrac{\partial^2\Phi}{\partial n^2}]$,
as there is no jump in the curvature.
Previous work has shown that the jump condition commutes with differentiation along the interface, $[\nabla^2_s\Phi]=\nabla^2_s[\Phi]$, see Ref. \cite{Xu2006}.
Also note that the jump in the Laplacian of the electric potential is zero, $[\nabla^2\Phi]=0$. 
It is thus possible to
write the jump in the second normal derivative as 
 \begin{equation}
	\biggl[\frac{\partial^2\Phi}{\partial n^2}\biggl]= -\nabla^2_s[\Phi] - H\biggl[\frac{\partial\Phi}{\partial n}\biggl],
	\label{eq:Phi_DoublePrime_simplified}
\end{equation}
where the expressions for $[\Phi]$ and $\left[\partial \Phi/\partial n\right]$ are already given in Eq. (\ref{eq:Phi_Difference}) and Eq. (\ref{eq:Phi_Prime_Jump}), respectively.

The jumps are calculated on the interface and extended to the surrounding nodes by extrapolating in the normal direction. The extended jumps 
at grid point, $\left[\Phi\right]_{gp}$, are given by
\begin{equation}
	\left[\Phi\right]_{gp}=\left[\Phi\right]+d\left[\frac{\partial\Phi}{\partial n}\right]+\frac{d}{2}\left[\frac{\partial^2\Phi}{\partial n^2}\right],
\end{equation}
where $d$ is the signed distance from the grid point to the interface. Using these extended jumps the corrections can be calculated 
in the Immersed Interface Method, see Ref. \cite{Lai2008} for the IIM implementation details.
 
\section{Numerical Implementation}
\label{sec:3.0}

The goal is to solve for the electric potential field in a rectangular domain discretized using a Cartesian grid with uniform grid spacing $h$, see Fig. \ref{fig:grid}.
Let the $\Omega^{min}$ domain contain the fluid with the \textit{smaller} electrical conductivity.
To proceed with the numerical implementation define the normal electric field in the $\Omega^{min}$ domain
as $r=\partial\Phi^{min}/\partial n$. Note that this quantity is only defined on the embedded interface.
Using a second-order time-discretization for the trans-membrane-potential results in
\begin{equation}
	C_m\frac{3 V_m^{n+1}-4 V_m^n+V_m^{n-1}}{2 \Delta t}+G_m V_m^{n+1}=-s^{min}r,
	\label{eq:time_vm}
\end{equation}
where $V_m^n$ and $V_m^{n-1}$ are the trans-membrane potentials in the two previous time-steps and are taken to be known 
while $r$ is the normal electric field at time $t^{n+1}$.
With this new definition and solving for $V_m^{n+1}$ the complete set of jump conditions can be rewritten as
\begin{align}
	[\Phi]&=\frac{1}{3C_m+2\Delta t G_m}(2\Delta t s^{min}r-4C_m V_m^n+C_m V_m^{n-1}), \label{JPhi} \\
	 [\frac{\partial\Phi}{\partial n}]&=-\frac{[s]}{s^{max}}r, \label{JNPhi} \\
	[\frac{\partial^2\Phi}{\partial n^2}]&=\nabla^2_s[\Phi]-H \left[\frac{\partial \Phi}{\partial n}\right], \label{JNNPhi}
\end{align}
where $s^{max}$ is the larger of the two fluid conductivities.
If $r$ is known then the jump conditions are fully defined. It would then be possible to use the Immersed Interface Method
to solve for $\Phi$ in the entire domain. 
Unfortunately, the value of $r$ is not explicitly known but must be determined as part of the problem.
Here a technique first introduced for the solution of the Stokes equations \cite{Li2007622} is used to determine $\Phi^{n+1}$ and $r$ 
simultaneously.

All the electric potential jump conditions are linear. Hence, all the IIM corrections 
will be also linear. The linear system which results from an IIM discretization
of the electric potential field equation can be written in an operator form as $\vec{L}\vec{\Phi}=\vec{C}$,
where $\vec{L}$ is the Laplacian operator and $\vec{C}$ is the vector containing the required corrections. 
The total correction $\vec{C}$ can be split into corrections due to $\vec{r}$ and the previous trans-membrane potentials, $\vec{V}_m^n$ and $\vec{V}_m^{n-1}$:
$\vec{C}=\vec{A_0}\vec{r}+\vec{B_0}$,
where $\vec{A_0}$ is a linear operator and $\vec{B_0}$ contains the known contribution from the previous voltages.
It is now possible to solve for the electric potential in the domain, $\vec{\Phi}=\vec{L^{-1}}\vec{A_0}\vec{r}+\vec{L^{-1}}\vec{B_0}$.
Let $\vec{M}$ be the one-sided normal derivative operator such that $\vec{M}\vec{\Phi}=\vec{r}$. 
Then $\vec{M}\vec{\Phi}=\vec{r}=\vec{M}\vec{L^{-1}}\vec{A_0}\vec{r}+\vec{M}\vec{L^{-1}}\vec{B_0}.$

This relation shows that the normal electric field, $\vec{r}$, has two linear contributions.
There is a contribution from the trans-membrane potentials at previous times and a contribution from the normal electric field itself. 
As the quantity $\vec{B}_0$ is known, that particular contribution can be explicitly calculated as
$\vec{r}_0=\vec{M}\vec{L^{-1}}\vec{B_0}$,
which is simply the solution of the electric potential field using only the contribution to the jump conditions from $\vec{V}_m^n$ and $\vec{V}_m^{n-1}$.
This electric potential solution is then 
projected onto the normal electric field space through the $\vec{M}$ operator.

The second contribution is from the still-unknown normal electric field, $\vec{r}$. This contribution, though, can be written as 
$\vec{M}\vec{L^{-1}}\vec{A_0}\vec{r}=\vec{A}\vec{r}$, where $\vec{A}\vec{r}$ is the solution of the electric potential projected onto the 
normal electric field space by only considering the $\vec{r}$ contributions to the jump conditions.

Using this simplified notation it can be stated that $\vec{r}$ is the solution to the following linear system:
$(\vec{A}-\vec{I})\vec{r}=-\vec{r_0}$.
As this linear system can not be written in explicit form, a matrix-free iterative linear system solution method is needed to obtain the solution. 
The quantity $\vec{r}$ is only defined on the interface and is thus a lower dimension than the computational domain.
Therefore a solver such as GMRES proves to be an excellent choice.

To complete this section a word needs to be said about computing
the normal electric potential derivative, $r=M\Phi$, and the calculation of surface Laplacian, $\nabla_s^2\left[\Phi\right]=-\nabla_s^2 V_m$,
at a point on the interface.
First consider the surface Laplacian of the trans-membrane potential. The trans-membrane voltage is only given on the interface. To 
facilitate calculations $V_m$ is extended in the normal direction into the embedding region near the interface. It has been shown 
that standard Cartesian derivatives equal surface derivatives if the quantity of interest is constant in the normal direction,
see the Closest Point Method for more details \cite{Macdonald2009}.

Next, let $I_3^\Phi$ be a bi-cubic (in 2D) or tri-cubic (in 3D) interpolant of the electric potential for the cell containing the interface point of interest, 
see Fig. \ref{fig:grid}. The normal derivative operator $M$ can be calculated as appropriate derivatives of the 
interpolant and the outward unit normal, $M:=\vec{n}\cdot\nabla I_3^\Phi$.
To calculate a normal derivative in a particular fluid it is simply necessary to
apply the corrections to the \textit{opposite} fluid's nodes, \textit{e.g.} if $r=\partial\Phi^+/\partial n$ the corrections would be 
applied to all nodes in the $\Omega^-$ fluid. In this way 
a particular fluid's normal electric field can be calculated and discontinuities in the field can be taken into account.

\section{The Numerical Algorithm}
	To determine the electric potential and trans-membrane potential at a time $t^{n+1}$ is assumed that the electric potential at the previous two time-steps are known:
	$V_m^n$ and $V_m^{n-1}$. The algorithm is then given as:
	\begin{description}
		\item[Step I:]{Solve for the electric potential field only using corrections due to $V_m^n$ and $V_m^{n-1}$: $\vec{\Phi_0}=\vec{L^{-1}}\vec{B_0}$ 
			using the given physical boundary conditions.}
		\item[Step II:]{Compute the constant contribution to the normal electric field as $\vec{r}_0=\vec{M}\vec{\Phi}_0$.}
		\item[Step III:]{Use a matrix-free iterative solver such as GMRES to solve \\ $\left(\vec{A}-\vec{I}\right)\vec{r}=-\vec{r}_0$. Each matrix-vector product
			$\left(\vec{A}-\vec{I}\right)\vec{r}$ requires the following steps:
			\begin{description}
				\item[Step 1:]{Solve for the electric potential using the given $\vec{r}$: $\vec{\Phi}_r=\vec{L}^{-1}\vec{A_0}\vec{r}$ using uniform boundary conditions 
					of $\vec{\Phi}_r|_{bc}=0$.}
				\item[Step 2:]{Calculate the normal electric field as $\vec{A}\vec{r} = \vec{M}\vec{\Phi}_r$.} 
				\item[Step 3:]{Return the quantity $\vec{A}\vec{r}-\vec{r}$ as the matrix-vector product.}
			\end{description}
		}
		\item[Step IV:]{The electric potential field in the computational domain is $\vec{\Phi}^{n+1}=\vec{\Phi}_0+\vec{\Phi}_r$. }
		\item[Step V:]{The new trans-membrane potential is updated using Eq. (\ref{eq:time_vm}).}	
	\end{description}

\section{Sample Result}
	Consider a spherical vesicle placed in an electric field in the absence of membrane conductivity, $G_m=0$. 
	In this simple case an analytic solution exists for the electric potential 
	and trans-membrane potential \cite{Schwalbe2011}. Let the electric field far from the vesicle be given by $\vec{E}^\infty=E_0\vec{\hat{y}}$.	
	The electric potential can be written in terms of spherical harmonics: $\Phi^\pm=-E_0 \left(Y_1^{-1}+Y_1^1\right)P^\pm$,
	where $Y_1^{\pm 1}$ are the first-order spherical harmonic modes and $P^\pm$ is a function of the membrane
	capacitance, $C_m$, the conductivity ratio between the inner and outer fluids, $\Lambda=s^-/s^+$, and time, $t$. 
	The trans-membrane potential has a solution of $V_m=\bar{V}(t)E_0\left(Y_1^{-1}+Y_1^1\right)$ where $\bar{V}(t)\sim 1-exp(-t)$.
	See Ref. \cite{Schwalbe2011} for details of solution.
	
	Using a conductivity ratio of $\Lambda=0.1$, membrane capacitance of $C_m=1$, and an external electric
	field strength of $E_0$=1 the time-evolution of the electric potential field and trans-membrane 
	potential in $\Omega=\Omega^-\cup\Omega^+$ for a spherical vesicles of radius 1 has been calculated up to a time of $t=20$. The domain
	spans the region $[-4,4]^3$ and Dirichlet boundary conditions are imposed on the computational domain boundary. 
	Convergence results for grid spacing ranging from $h=0.0313$ to $h=0.125$ using a time step of $\Delta t=h$ are reported 
	in Table \ref{table:errorTable}. The electric potential, trans-membrane potential, and normal electric field ($r$ in the numerical method)
	are all consistently second-order accurate in the $L_\infty$-norm error.
	
\begin{table}[h]
    \scriptsize
	\centering	
	\caption{ Convergence results for the electric potential, trans-membrane potential, and electric field normal to the interface
	 for a spherical vesicle of radius 1. The normalized inner fluid conductivity is 0.1 while the outer fluid conductivity is set to 1. The 
	 membrane capacitance is set to $C_m=1$ while the conductivity is $G_m=0$. The external electric field has strength of $E_0=1$.
		The vesicle is placed in a $[-4,4]^3$ domain while the time step is fixed as $\Delta t=h$. All errors are computed at a time of $t=20$.
		Comparison is done against analytical results of Schwalbe \textit {et al.}  \cite{Schwalbe2011}}\vspace{4pt}
	\label{table:errorTable}	
	\begin{tabular}{| c | c c | c c | c c |}
	\hline
	  & \multicolumn{2}{ |c| }{ Electric Potential} & \multicolumn{2}{ |c| }{ Trans-Membrane Potential} & \multicolumn{2}{ |c|}{Normal Electric Field} \\ 
	\hline
		$h$ & $L_\infty$ & Order & $L_\infty$ & Order & $L_\infty$ & Order  \\ \hline
		0.1250 & $4.5134\times 10^{-3}$ & - & $3.9191\times 10^{-3}$ & - & $8.8146\times 10^{-4}$ & - \\ \hline
		0.0833 & $2.0335\times 10^{-3}$ & 1.89 & $1.7681\times 10^{-3}$ & 1.93 & $3.8786\times 10^{-4}$ & 2.4 \\ \hline
		0.0625 & $1.1690\times 10^{-3}$ & 1.82 & $1.0209\times 10^{-3}$ & 1.86 & $2.0215\times 10^{-4}$ & 2.22 \\ \hline
		0.0417 & $5.1347\times 10^{-4}$ & 1.89 & $4.4380\times 10^{-4}$ & 1.93 & $8.6623\times 10^{-5}$ & 2.34 \\ \hline
		0.0313 & $3.0121\times 10^{-4}$ & 1.85 & $2.6004\times 10^{-4}$ & 1.88 & $5.1190\times 10^{-5}$ & 2.25 \\ 
	\hline	
	\end{tabular}
\end{table} 

\begin{figure}
	
	\centering	
	\begin{minipage}{0.3\textwidth}
		\centering
		\includegraphics[width=1.75in]{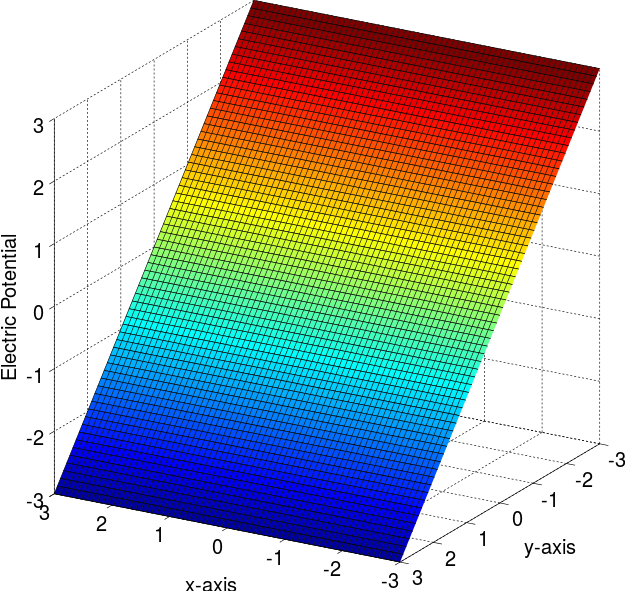} \\[5ex]
		\includegraphics[width=1.75in]{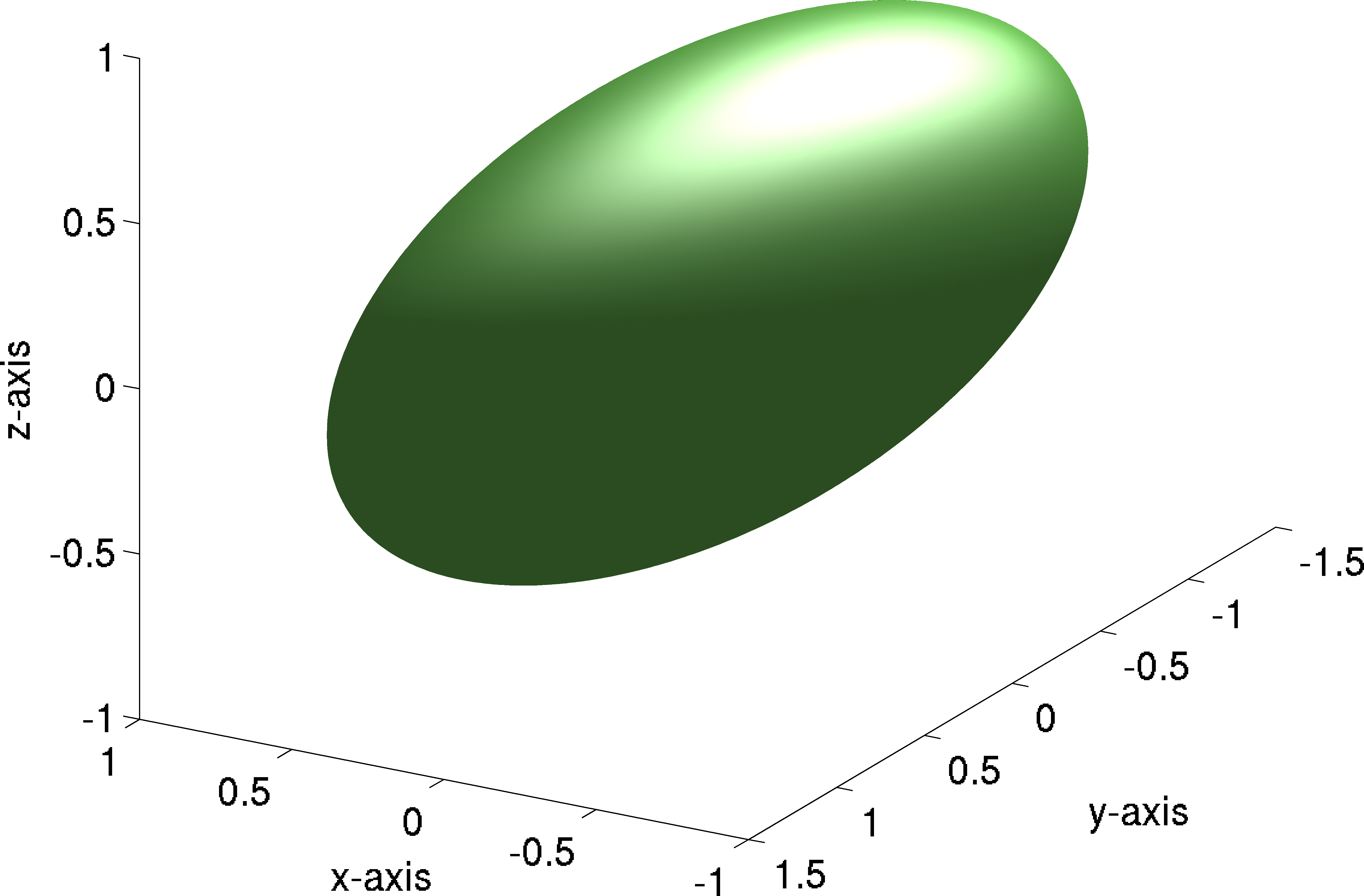} \\
		t=0
	\end{minipage} \hfill
	\begin{minipage}{0.3\textwidth}
	\centering
		\includegraphics[width=1.75in]{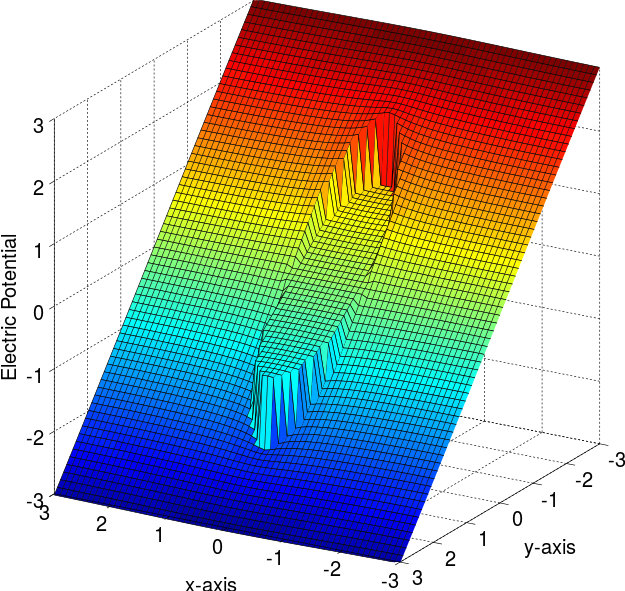} \\[5ex]
		\includegraphics[width=1.75in]{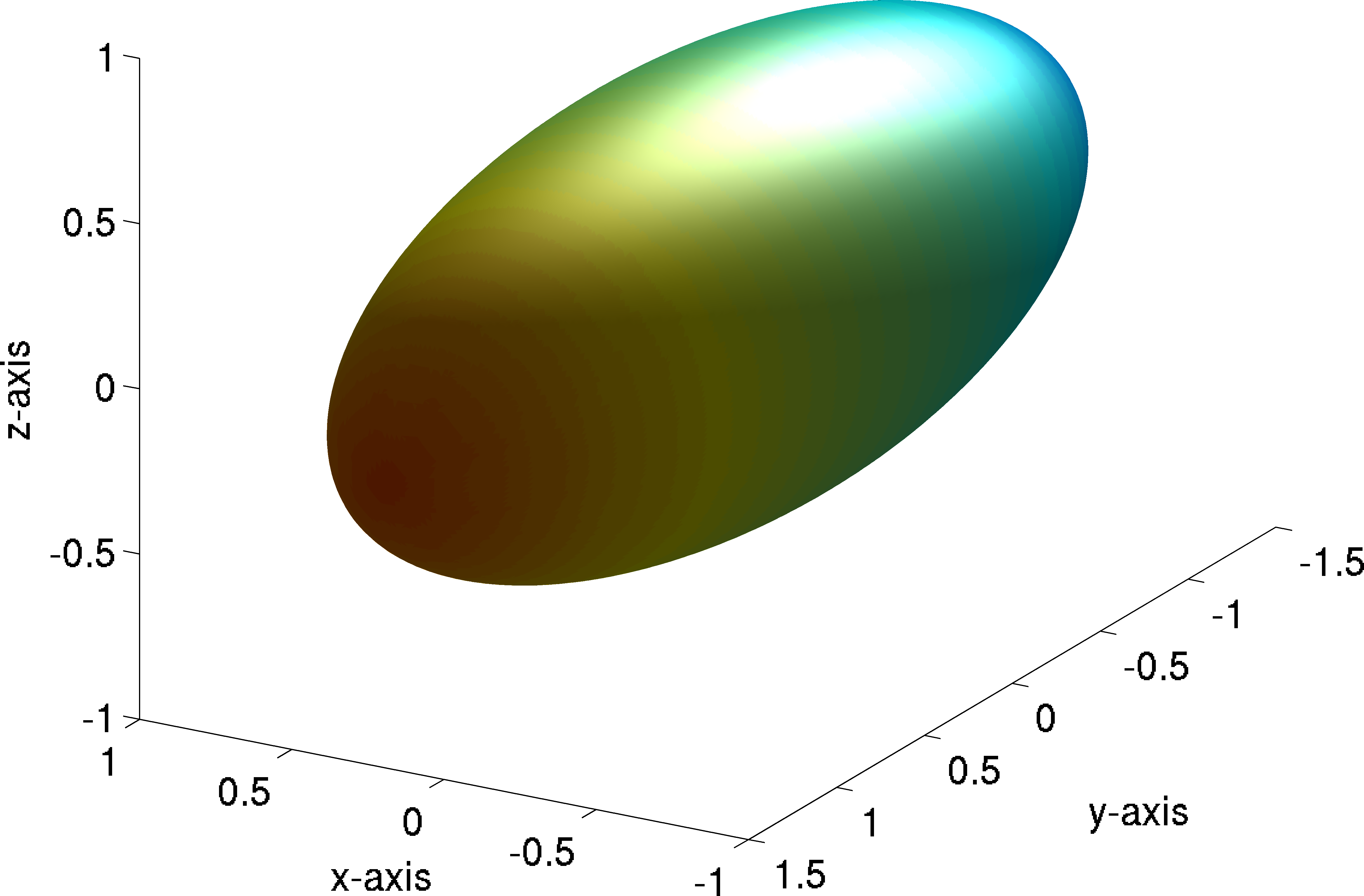} \\
		t=46.875
	\end{minipage} \hfill
	\begin{minipage}{0.3\textwidth}
	\centering
		\includegraphics[width=1.75in]{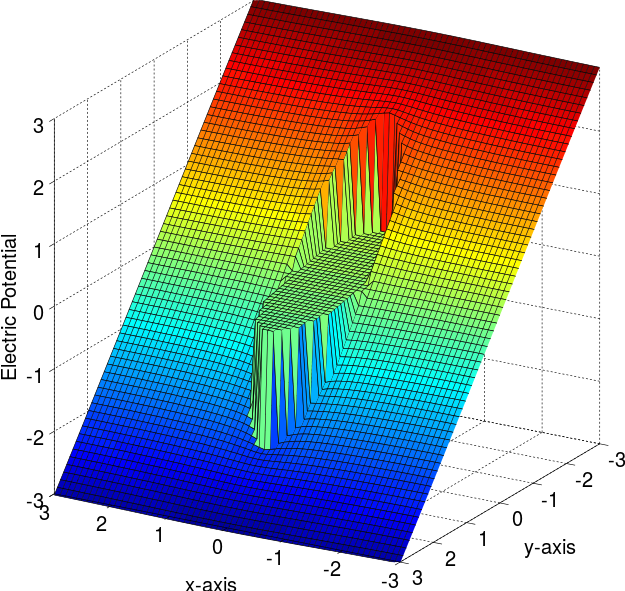} \\[5ex]
		\includegraphics[width=1.75in]{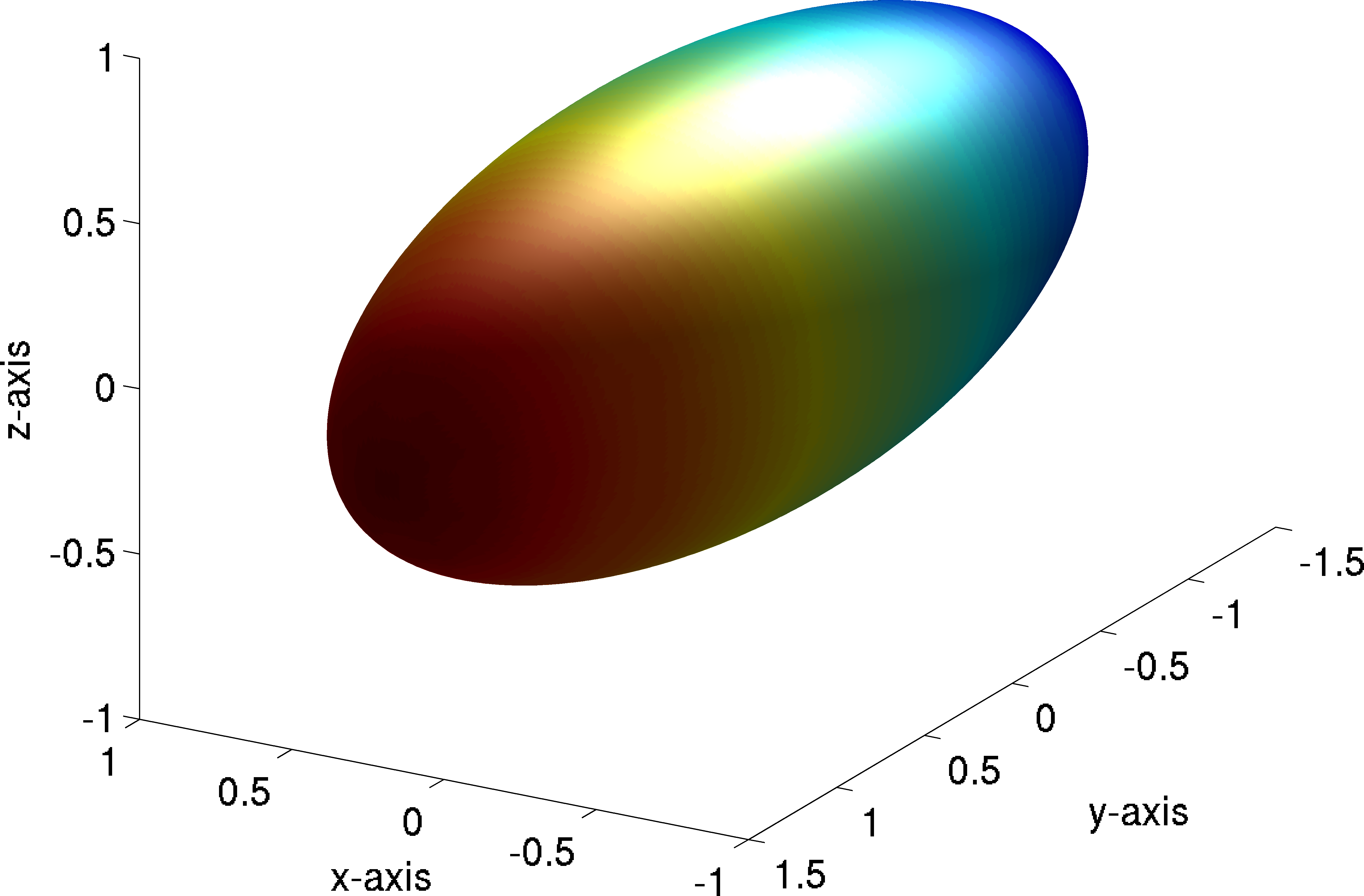} \\
		t=187.5
	\end{minipage}
	\caption{Sample results for an elliptic interface in an electric field. The top row are 
		for the $z=0$ plane while the bottom row is the trans-membrane potential. The trans-membrane potential
		has values between -2 (blue) to +2 (red).}
	\label{fig:elliptic}		
\end{figure}

	As an example of solving the system for a non-spherical shape consider an ellipsoidal shape with an axis length of 3.7 in the $\hat{y}$-direction
	and an axis lengths of 1.33 in the $\hat{x}$- and $\hat{z}$-directions. The electric field is in the $\hat{y}$-direction and has a far-field strength
	of 1. The membrane capacitance is set to $C_m=1$ while the conductivity is set to a small, but non-zero value, $G_m=0.001$. In this case the inner
	fluid conductivity is set to $s^-=0.05$ while the outer conductivity is unity, $s^+=1$. The time-evolution of the electric potential on the $z=0$ plane
	and the evolution of the trans-membrane potential are shown in Fig. \ref{fig:elliptic}. Over time the trans-membrane saturates between values of -2 and +2
	and the potential of the inner fluid flattens out, which matches what is expected for vesicles \cite{Schwalbe2011}.
	
\section{Conclusion}
	In this letter a method to model the trans-membrane potential and the surrounding electric potential for a vesicle exposed 
	to an electric field has been developed. The jump conditions depend on the electric field normal to the membrane, which is 
	determined as part of the solution. Overall the method demonstrates second-order accuracy.
	
	This is part of an ongoing work to investigate the electrohydrodynamics of lipid bilayer vesicles. In the future the 
	electric potential solver will be coupled to a general multiphase solver to investigate the dynamics of vesicles 
	in the presence of electric fields.
	

 \bibliography{SISCELEC091513}

\end{document}